\documentclass[useAMS,usenatbib]{mn2e}
\usepackage{graphicx}

\title[BS\,196: an old SMC star cluster]{BS\,196: an old  star cluster  far
  from  the SMC main body\thanks{Based on observations made with the SOAR 
4.1-m telescope (proposal SO2007B-007)}}
\author[E. Bica, J. F. C. Santos Jr. and A. A. Schmidt]
{E. Bica$^{1}$\thanks{E-mail: bica@if.ufrgs.br}, 
J. F. C. Santos Jr.$^{2}$ and A. A. Schmidt$^{3}$\\
$^{1}$Departamento de Astronomia, Universidade Federal do Rio Grande do Sul,
  Av. Bento Gon\c calves 9500, Porto Alegre 91501-970, RS, Brazil\\
$^{2}$Departamento de F\'{\i}sica, ICEx, Universidade Federal de Minas Gerais,
  Av. Ant\^onio Carlos 6627, Belo Horizonte 31270-901, MG, Brazil\\
$^{3}$Departamento de Matem\'atica, Universidade Federal de Santa
  Maria, Av. Roraima 1000, Santa Maria 97105-900, RS, Brazil}

\begin{document}

\date{Accepted 2008 September 10.  Received 2008 August 26; in original form
  2008 July 3}

\pagerange{\pageref{firstpage}--\pageref{lastpage}} \pubyear{}

\maketitle

\label{firstpage}

\begin{abstract}
  We present {\it B} and {\it V} photometry of the outlying SMC star 
cluster BS\,196 
with the 4.1-m SOAR telescope. The photometry is deep (to {\it V}$\approx$25) 
showing $\approx$3 mag below the cluster turnoff point (TO) 
at {\it M$_{\it V}$}=2.5 
(1.03\,M$_{\odot}$). The cluster is located at the SMC distance.
The colour-magnitude diagram (CMD) and isochrone fittings 
provide a cluster age of 
5.0$\pm$0.5\,Gyr, indicating that this is one of the 12  oldest clusters 
so far detected in the SMC. The 
estimated metallicity  is [Fe/H]$=-1.68\pm0.10$.
The structural analysis gives by means of King profile fittings a core radius 
{\it R}$_{\rm c}$=8.7$\pm$1.1\,arcsec 
(2.66$\pm$0.14\,pc) and a tidal radius {\it R}$_{\rm t}$=69.4$\pm$1.7\,arcsec 
(21.2$\pm$1.2\,pc). BS\,196 is rather loose with a concentration 
parameter {\it c}=0.90. With {\it M}$_{\it V}$=$-1.89\pm0.39$, BS\,196  
belongs to the class of  intrinsically fainter  SMC clusters, as compared 
to the well-known populous ones, which starts to be explored.

\end{abstract}


\section{INTRODUCTION}
Luminous old SMC star clusters (we will refer as such to the red SMC 
clusters with 
age $>$ 1\,Gyr -- including intermediate age clusters (IAC) to 
globular clusters)
have been systematically observed  in the last decade,  using mainly 
{\it CT$_1$} 
photometry at CTIO  \citep[e.g.][]{psc01,psg07}, and
Hubble photometry with different cameras \citep[e.g.][]{msf98,ggg08}.
The age 
and age-metallicity distributions in these studies reveal essential 
features of the star formation history of the
SMC system. Any new star cluster added to these distributions provides  
fundamental pieces of information to them.
 
The star cluster BS\,196 was discovered on Sky Survey plates and  first 
reported by \cite{bs95}. It was also listed
by \cite{bd00} in a update of the SMC catalogue. Recently, 
two star clusters from  Bica \& Schmitt's  catalogue were found  to be  
old SMC clusters. The 4.3\,Gyr old \citep{ssn07} cluster BS\,90 is 
unique, being  projected inside the large H\,II region complex NGC\,346 
and BS\,121 with 2.3\,Gyr \citep{psg05} is projected on  the SMC 
main body.
 
BS\,196 appears  to be  a less populous cluster on the Sky Survey plates, 
with a diameter somewhat smaller than 1\,arcmin.  
It is  located at $\alpha_{2000}$ = 1h48m02s and $\delta_{2000} = 
-70^\circ$00\arcmin15\arcsec. It is 
projected  5.1$^\circ$ to the Northeast  of  the SMC bar and the optical 
centre at  $\alpha_{2000}$=00h52m45s,
$\delta_{2000}=-72^\circ$49\arcmin43\arcsec~ \citep{csp01}, being one 
of the angularly farthest SMC clusters known \citep{bs95}. It is more 
distant from the bar than Lindsay\,1, an extreme cluster to the West, together 
with AM-3  \citep{bs95}.  It is fundamental to carry out deep photometry 
of the cluster  BS\,196 to unveil its nature, whether associated to the SMC 
or a far globular cluster in the Galactic halo.
 
This paper is organized as follows. In Sect. 2 {\it BV} photometry of BS\,196 
obtained with the SOAR telescope is presented. In Sect. 3  we determine 
the cluster astrophysical 
parameters by means of  Padova isochrone fittings, finding that it 
belongs to the SMC. In Sect. 4, cluster structural properties are derived. 
In Sect. 5 we discuss  the cluster age and 
metallicity  in the context of the SMC  enrichment and star formation 
histories.  
Concluding remarks  are given in Sect. 6.
 
\section{OBSERVATIONS}

The SOAR optical imager (SOI) mounted in a bent-Cassegrain configuration 
to the 4.1-m SOAR telescope (Cerro Pach\'on, Chile) was employed 
to observe BS\,196 in service mode. Images taken with Bessel {\it BV} filters 
were obtained on
the photometric night of 2007 November 11, using the SOI mini-mosaic of 
two E2V 2x4k 
CCDs (1 pixel = 15$\mu$m) to cover a 5.2\,arcmin square field of view. 
The observations
were carried out with the CCDs binned to 2x2 pixels to yield a scale of 
0.154\,arcsec/pixel. The readout mode was set to {\it fast} (readout noise
4.1\,e$^-$) and the gain was 1.9 e$^-$ADU$^{-1}$. The seeing FWHM was 
$\sim$0.95\,arcsec in {\it B} and $\sim$0.8\,arcsec in {\it V} with the 
observations being taken at airmass 1.31. Two images in each filter
were obtained, with single exposure times of 480\,s in {\it B} and 195\,s in 
{\it V}.
The telescope pointing was so as to carefully center the object on one of the 
CCDs in order to avoid the gap of 7.8\,arcsec between them.
The combined {\it V} image of BS\,196 is shown in Fig. 1, as an extraction
of the SOI frame. 

\begin{figure}
\centering
\includegraphics[width=7.8cm]{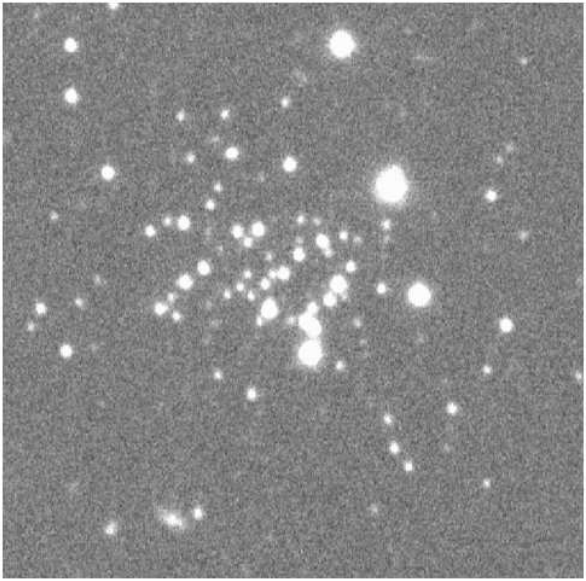}
 \caption{SOI {\it V} image extraction of BS\,196 field. 
North is up and East to the left. The field dimensions are 
$1.2\times1.2$ arcmin$^2$.}
\end{figure}

Standard fields were also observed with airmasses bracketing the target
airmass for photometric calibration. The NGC\,121 Stetson's photometric 
standard field\footnote{http://www3.cadc--ccda.hia--iha.nrc--\\cnrc.gc.ca/community/STETSON/standards} \citep{s00} provides {\it B} and {\it V} 
magnitudes of $\approx$110 stars. 
Since it covers a larger field than the SOI FOV, a fraction of it containing
28 stars was observed twice, with different airmasses, on the same night.
Also the standard stars A, C and D in the T\,Phe field were observed once. 
 
The CCD frames were reduced with {\sc iraf} software \citep{t93}.
They were pre-reduced (bias/overscan subtraction, 
flat-field division using dome flats, trimming of overscan frame regions 
and  transformation of MEF files into single FITS files) 
using SOAR scripts suited to operate on SOI images. 
Removal of cosmic rays was performed by combining the two exposures
in each filter. 

The combined {\it B} image was registered using selected star coordinates 
of the combined {\it V} image. Then a single list of star coordinates was
generated (using {\it V} image) for all stars brighter than 5 times the sky 
fluctuation and their aperture photometry computed. As in the previous steps,
{\sc daophot} as implemented in {\sc iraf} was employed to perform 
point spread function (PSF) photometry. For the target cluster 
there was no 
particular difficulty in finding relatively bright, isolated stars for 
building the PSF model because of its location far from the SMC main body. 
Three stars with brightness covering those of cluster stars were chosen as 
PSF stars in each {\it B} and {\it V} frames.
The instrumental magnitudes were obtained by running {\sc daophot} task
{\sc allstar}. Besides those, {\sc allstar} also outputs photometric 
quality parameters, namely, $\chi$, which provides the goodness of PSF fitting
(ideally equal to 1), and $sharp$, which accounts for the difference between 
the observed width of an object and the width of the PSF model (ideally equal
to zero).  Only stars with $\chi<3.0$ and -0.5$<{sharp}<$0.5  
were kept. The final table of magnitudes contains
660 stars. Aperture corrections 
of -0.346 and -0.343 for {\it B} and {\it V} respectively were applied
to the instrumental magnitudes in order to bring those to the standard
stars scale. 

Given the small range of airmasses covered by the standards we did not 
derive extinction coefficients but use k$_{\it B}=0.25$ and k$_{\it V}=0.14$ 
(CTIO average) as constants in the transformation equations.
The photometric calibration was accomplished by using {\sc fitparams iraf} 
task with the following transformation equations:

\begin{equation}
mb=B+b_1+b_2(X_B-1.4)+b_3(B-V)+b_4(B-V)X_B
\end{equation}

\begin{equation}
mv=V+v_1+v_2(X_V-1.4)+v_3(B-V)+v_4(B-V)X_V
\end{equation}

\noindent where $b_i$ and $v_i$ ($i=1,...,4$) are the derived coefficients,
$mb$ and $mv$ are the instrumental magnitudes, $B$ and $V$ are 
the standard magnitudes and $X_B$ and $X_V$ the airmasses.
Note that $b_2$=k$_{\it B}$ and $v_2$=k$_{\it V}$ and the numerical factor
1.4 were used to minimize the significance of the extinction coefficients.
Also the initial zeropoint magnitude was 25.0.
The fittings rms and derived coefficients are given in Table~1.
 
\begin{table}
 \centering
  \caption{RMS and coefficients of transformation equations.}
  \begin{tabular}{crr}
  \hline
   $i$    & {\it B}   & {\it V}\\
 \hline
    1     &  -0.230$\pm$0.016 &  -0.3065$\pm$0.0057\\
    2     &   0.25       &   0.14\\
    3     &  -0.480$\pm$0.099 &  -0.384$\pm$0.046\\
    4     &   0.357$\pm$0.074 &   0.376$\pm$0.034\\
   RMS    & 0.026&                  0.017            \\
\hline
\end{tabular}
\end{table}

The transformation equations were then inverted and applied to the 
target cluster. 

Photometric errors in the {\it B} and {\it V} bands are shown in 
Fig. 2. As a consequence of the deeper image in {\it B}, the errors are 
smaller in {\it B} than in {\it V} at a same magnitude. At mag=24 
$\sigma_B\approx$0.05  and  $\sigma_V\approx$0.10.

\begin{figure}
\includegraphics[width=9cm]{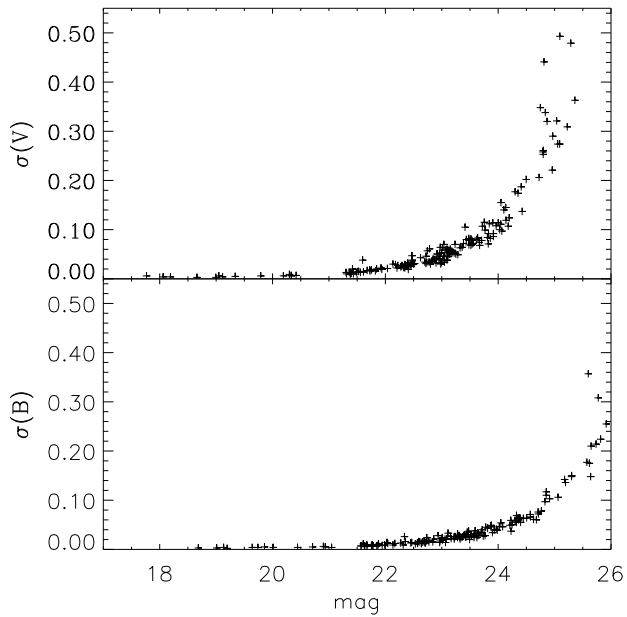}
 \caption{Photometric errors in {\it B} (lower panel) and {\it V} 
(upper panel) as a function
of the respective magnitude. The data are limited to errors smaller than 0.5
and cover an area about 4 
times the cluster visual area ({\it r} = 130\,pix), centred in the cluster.}
\end{figure}

\section{COLOUR-MAGNITUDE DIAGRAM AND ASTROPHYSICAL PARAMETERS}

We indicate  in Fig. 3 a series of concentric spatial regions selected for 
building {\it V} $\times$ ({\it B -- V}) CMDs. 

\begin{figure}
\includegraphics[width=8cm]{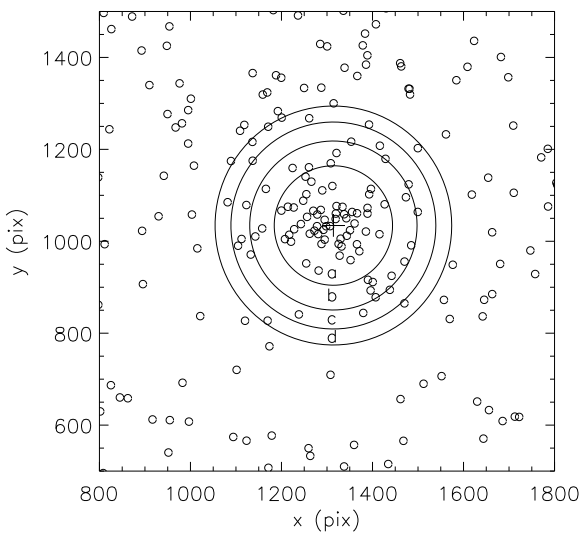}
 \caption{Spatial distribution of stars in the field of BS\,196, 
where the
cluster centre (plus symbol) and the annular regions used to build the 
CMDs of Fig. 4 are shown (a, b, c and d). The field size is 
150$\times$150 arcsec$^2$. North is up and East to the left.}
\end{figure}

The CMDs are shown in Fig. 4. The central part 
({\it r}\,$ < $\,20\,arcsec) shows a clear main sequence (MS) 
and giant branch (GB). In the outer  rings to as far as 
{\it r}\,$ < $\,40\,arcsec the MS can still be traced.

\begin{figure}
\includegraphics[width=9cm]{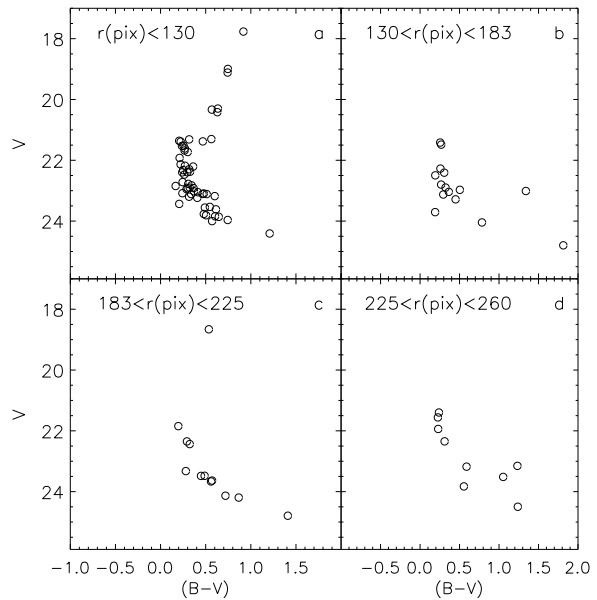}
 \caption{CMDs of annular regions of same area centred in the cluster. 
The circular inner region has the cluster visual radius (a). The boundary
radius, in pixels, of outer adjacent rings are indicated (b, c, d).
The panels contain data with photometric errors less than 0.5 mag.}
\end{figure}

 Before fitting isochrones to the cluster inner region CMD,
we searched the literature for values of reddening and distance modulus
aiming at constraining their ranges in the process. 
The reddening was taken from H\,I \citep{bh82}
and dust \citep{sfd98} maps. For the cluster galactic coordinates 
($l=295.98^{\circ}$, $b=-46.35^{\circ}$) both maps agree:
{\it E}({\it B -- V})$=0.02\pm0.01$ \citep{bh82} and 
{\it E}({\it B -- V})$=0.024\pm0.004$ \citep{sfd98}, where the first 
value has been brought to the scale of \cite{sfd98}. We adopt the 
mean value {\it E}({\it B -- V})$=0.02\pm0.01$.
The distance modulus was allowed to change
within ({\it m-M})$_{\circ}=18.9\pm0.2$ \citep{w97}, where 
the uncertainty accounts for the SMC line-of-sight depth \citep{csp01}.

Fortunately the 
cluster presents a well populated TO and well defined giant (RGB) and 
subgiant (SGB) branches owing to the
accurate photometry in this uncontaminated field, which helps 
further to narrow down the uncertainties associated with isochrone fitting.
Due to the presence of giants, we do not identify any degeneracy yielded
by combinations of reddening, distance modulus, age and metallicity since
each of these parameters have a different effect on the isochrones.
Considering the domain of ages dealt with in the present study, the main 
parameters shaping an isochrone are its age (affects principally
the TO) and its metallicity (affects principally the RGB but also the TO). 
On the other hand, reddening and distance modulus are related to displacement 
of data and isochrones in the CMD plane. A certain
combination of reddening and distance modulus applied to the data matches
them with a selected isochrone in the CMD plane.   

The procedure employed to find the best representative 
isochrones of the cluster CMD was interactive:

(i)~ we used data of the inner region of the cluster ($r<20$\,arcsec)
with photometric errors below 0.1 mag. 

(ii)~the initial values {\it E}({\it B -- V})$=0.02$
and ({\it m-M})$_{\circ}=18.9$, as justified above, were then applied to the
data.

(iii)~we then visually chose Padova isochrones \citep{gbb02} which best 
matched the corrected data in the CMD.
A good match was achieved by using the isochrone with log\,{\it t} = 9.7
and Z=0.0003, by revising to ({\it m-M})$_{\circ}=18.95$ the distance 
modulus correction applied to the data and by keeping the initial 
reddening, {\it E}({\it B -- V})$=0.02$.
Although this distance modulus places the cluster slightly farther than that
derived from a series of clusters \citep{csp01}, namely 
({\it m-M})$_\circ$=18.77, we conclude that BS\,196 appears to be 
located at the SMC distance.

(iv)~in order to account for the uncertainties in age, additional isochrones
of  log\,{\it t} = 9.6, 9.8 and Z=0.0003 were superimposed in the CMD
to the data corrected for the extreme and average values that reddening 
and distance modulus may assume due to errors of 0.01 mag and 0.05 mag, 
respectively. Fig. 5 shows this step.
Considering that the three isochrones encompass the data, producing 
reasonable fits whenever the errors in {\it E}({\it B -- V}) and 
({\it m-M})$_{\circ}$ are taken into account,  the adopted age is the 
average log\,{\it t}$=9.70\pm0.04$ (5.0$\pm$0.5\, Gyr), 
which comes from assigning weight two to the central isochrone and weight one
to the marginal ones. 

\begin{figure}
\includegraphics[width=9cm]{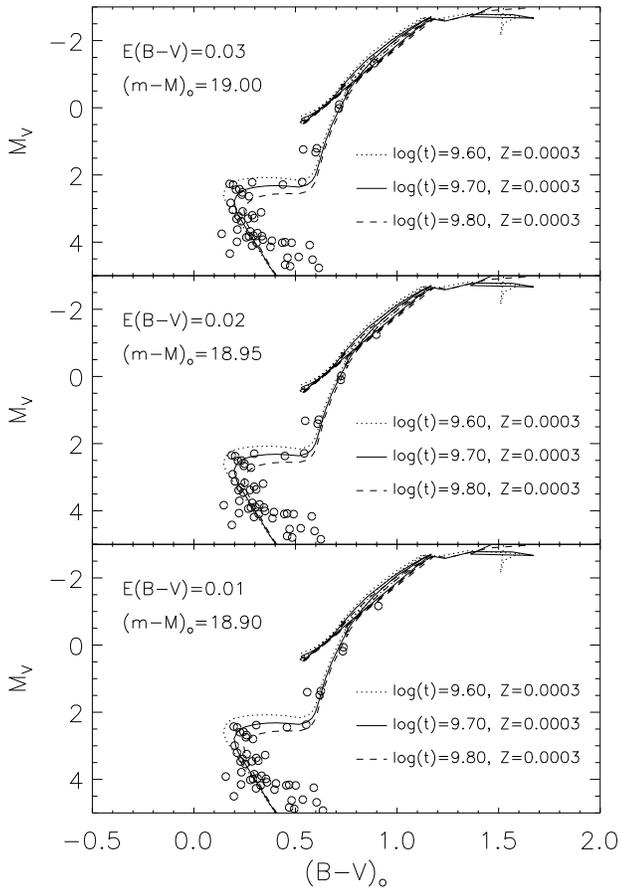}
 \caption{Isochrone fitting to the cluster CMD inner region. Stars with 
photometric errors above 0.1 mag have been excluded. The three panels show 
the effect of different reddening and distance modulus applied to the data.
Each panel contains the same set of Padova isochrones of Z=0.0003
and different ages.}
\end{figure}

(v)~the cluster metallicity was derived by
matching isochrones of age log\,{\it t}$=9.70$ and different
metallicities to the data which was corrected for  
({\it m-M})$_{\circ}=18.95$ and {\it E}({\it B -- V})$=0.02$.
These fittings are shown in Fig. 6 for isochrones with Z=0.0001, 0.0003, 
0.0005, 0.0007. They provide the stellar mass in the TO region: 
1.03\,M$_{\odot}$. 
In a procedure similar to that used for age 
determination, an average metallicity of Z=$0.00040\pm0.00009$
([Fe/H]=$-1.68\pm0.10$) was derived by assigning weight two to the 
central two isochrones (Z=0.0003 and Z=0.0005), given that both
seem to represent equally well the overall stars loci in the CMD
and weight one was assigned to the marginal isochrones (Z=0.0001 and 
Z=0.0007), which encompass most of the stars brighter than
1 mag below the TO.

\begin{figure}
\includegraphics[width=8.5cm]{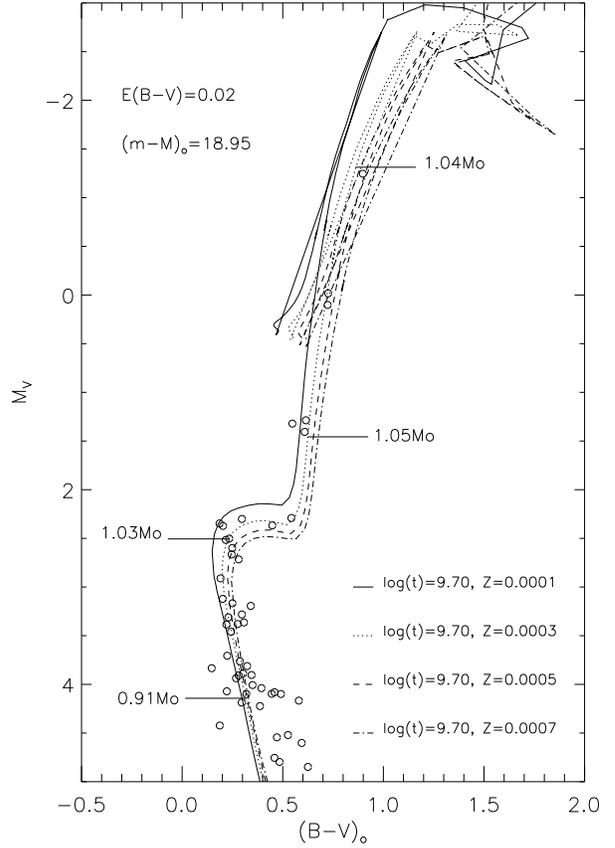}
 \caption{Isochrone fitting to the cluster CMD inner region. Stars with 
photometric errors above 0.1 mag have been excluded. The data have been 
corrected by the best achieved values of reddening and distance modulus.
The effect of different isochrone metallicities for
the best fitted age is shown. Stellar masses in specific isochrone loci are 
also indicated.}
\end{figure}

 The Padova isochrones have a widespread use and are successful 
in reproducing the stars loci in CMDs of stellar
clusters found in nature. However, debate on specific details of 
stellar evolution (e. g. convective overshooting) have not been 
settled to a concensus among the groups that build the stellar 
evolutionary models. To evaluate the effect of using different sets of
isochrones on the astrophysical parameters derived for BS\,196,
we compare in Fig. 7 Padova isochrones with
those of Y$^2$ \citep{y01} of similar properties. 
Fig. 7a compares  isochrones of Z=0.0003 and different ages, showing
an overall resemblance for similar age isochrones. Slight differences 
occur in the TO region in the sense that Y$^2$
isochrones of same age are brighter and bluer than the Padova isochrones,
the effect being less relevant for older ages. Fig. 7b presents isochrones
of log\,{\it t}$=9.70$ and different metallicities, again showing the
overall similarities between sets.  Even slighter loci differences are 
seen as an effect of metallicity as compared to that on age. The small
differences are in the sense that  Y$^2$
isochrones  are brighter and bluer than the Padova isochrones of same 
metallicity. All these slight isochrone differences are within our errors
determined for the cluster parameters. 
In summary, the close resemblance
of these sets for ages around 5\,Gyr indicates that our estimates of age and 
metallicity for BS\,196 does not depend on whether we choose one or another
set. It is worth mentioning on this regard that Padova and Y$^2$ isochrones
are based on stellar evolutionary tracks built with different input physics
and different prescriptions for conversion from the theoretical 
to the observational plane.

\begin{figure}
\includegraphics[width=8.5cm]{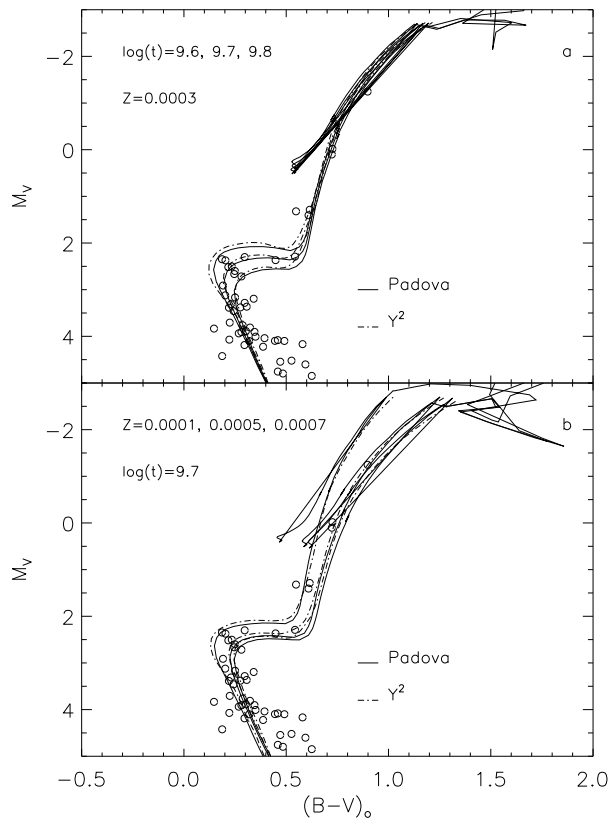}
 \caption{Comparison of Padova and Y$^2$ isochrones for the interval
of age and metallicity investigated in the present work. The data is
the same of Fig. 6.
(a)~Effect of changing age at Z=0.0003; (b)~Effect of changing metallicity
at log\,{\it t}$=9.70$.}
\end{figure}

\section{STELLAR DENSITY PROFILE AND STRUCTURAL PARAMETERS}

The first step towards the construction of a representative stellar density
profile is the determination of the object centre, which was acomplished 
by the moment-based ellipse fitting, as described by \cite{bds95}, to the 
images of the cluster. The cluster overall structure is well represented
by an ellipse with centre at $x= 1314.05\pm0.34$,
$y = 1034.34\pm0.24$ (pixel coordinates, see Fig. 3), ellipticity 
$\epsilon=1-b/a = 0.27\pm0.05$ and position angle 
$PA = (-3.07\pm0.61)^{\circ}$.

To take into account the cluster projected elliptical shape, the radial 
density profile (RDP) was computed within circular rings but with the 
star coordinates
modified by including the dependence on their position angles relative 
to the ellipse orientation $PA$ and the cluster $\epsilon$. The procedure
is similar to a deprojection of the cluster on its major axis.
 
We employ two-parameter (central stellar surface density, $\sigma_0$, and 
core radius, {\it R}$_{\rm c}$) and three-parameter ($\sigma_0$, 
{\it R}$_{\rm c}$ and tidal radius, {\it R}$_{\rm t}$) King profiles 
\citep{k62,k66} to derive 
the cluster structural properties. Due to the non-populous nature of the 
cluster we do not use  integrated fluxes, rather we employ star counts to 
derive radial density profiles.
Besides the cluster centre given above, which corresponds to the
intensity-weighted centroid, we tested a few other centers 
to optimize higher central counts, but the results were essentially the 
same.

   The results of the analysis are given in  Fig. 8. The cluster has a 
well defined profile (Fig. 8a), which indicates a limiting
radius of 47$\pm$2\,arcsec ($14.38\pm0.81$\,pc), determined where profile 
and background merge 
(Fig. 8b). The background has been determined by fitting 
a constant to the outermost 2 rings. In Fig. 8b this background is 
shown sided by its 1\,$\sigma$ dispersion.
The fitted constant background was then subtracted from the overall
surface density and a King-profile fitting was performed.

\begin{figure}
\includegraphics[width=8.5cm]{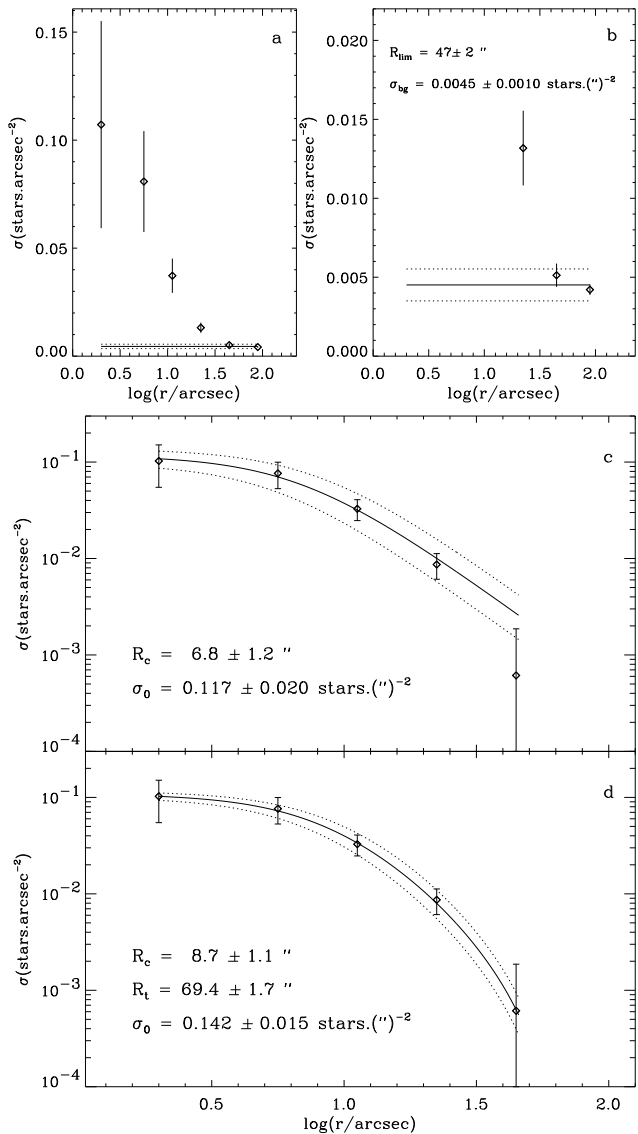}
 \caption{Radial density profile analysis of BS\,196. (a) The RDP 
with poissonian error bars; (b) Zoomed version of (a) where it is 
determined the cluster limiting size and the constant background; 
The RDP in log scale with  2-parameter (c) and 3-parameter (d)
King profile fittings for which central density, core and tidal radii
are indicated. 1\,$\sigma$ dispersion for the fittings are indicated 
by dotted lines.}
\end{figure}

The fitted functions are 
presented in Fig. 8c and 8d in log scales together with 
the best fit parameters. 
The two-parametric King function should better represent the cluster inner
regions and the three-parametric King function should provide a better estimate
of the cluster overall structure. In both fittings the estimates of 
{\it R}$_{\rm c}$ agree within the uncertainties. The two-parametric King 
function gives a slightly lower value of $\sigma_0$ than the three-parametric 
one.
The tidal radius is well constrained in spite of the fluctuations in the
density of cluster stars in its outskirts, almost at the background level.
Such fluctuations (represented by poissonian errors in Fig. 7) 
are taken into account in the fitting by applying a weigthed 
least-squares method.   

Based on the three-parametric King function fitting, the structural
parameters of BS\,196 are $\sigma_0$=0.142$\pm$0.015\,stars.arcsec$^{-2}$ 
(1.52$\pm$0.16\,pc$^{-2}$), 
{\it R}$_{\rm c}$=8.7$\pm$1.1\,arcsec 
(2.66$\pm$0.14\,pc) and a tidal radius {\it R}$_{\rm t}$=69.4$\pm$1.7\,arcsec 
(21.2$\pm$1.2\,pc). BS\,196 is rather loose 
with a concentration parameter {\it c}=0.90.

In general the low mass stars do not contribute significantly to the 
cluster integrated light \citep{sbd90}.
Thus the  observed stars in BS\,196 lead to an estimate of the cluster 
integrated magnitude of {\it M}$_{\it V}$=$-1.89\pm0.39$, and integrated 
colour ({\it B -- V})=$0.563\pm0.090$. The uncertainties account for an error
of 10 per cent in the cluster visual radius.  

The cluster is less luminous than AM-3 with 
{\it M}$_{\it V}$=$-3.5\pm0.5$ \citep{dc99}.
Such SMC clusters are intrinsically fainter than  the populous ones in 
both Clouds \citep[e.g.][]{h60}, and resemble low-mass Palomar 
globular clusters in the Milky Way \citep{bb08}.

\section{THE CLUSTER IN THE CONTEXT OF SMC EVOLUTIONARY HISTORY}

In this section we compare properties of BS\,196 with other SMC clusters of 
similar age. \cite{r00} found evidence of two coeval groups of SMC clusters 
centred at 2 and 8\,Gyr. BS\,196 is located midway between the two groups. 

Old SMC clusters in the following are defined as those older than the lower 
limit of the LMC age gap or $\approx$4\,Gyr \citep{psg02}.  We indicate 
in Table 2  the  genuine (age $^>_{-}$ 9\,Gyr) globular clusters in the SMC 
and intermediate age clusters (IACs) in the range  4-9\,Gyr. There are 
15 objects including the present one. Ages, metallicities and references 
for these clusters are given
in Table 2, together with equatorial coordinates and designations 
\citep{b08}. The  BS\,196 properties are included.  

\begin{table*}
 \centering
 \begin{minipage}{150mm}
  \caption{SMC clusters older than 4\,Gyr.}
  \begin{tabular}{lrrrrl}
  \hline
Designations  &$\alpha_{2000}$ &$\delta_{2000}$ & age & [Fe/H] & References\\
              & (h:m:s)        &  ($^{\circ}:\arcmin:\arcsec$) & (Gyr) & & \\
 \hline
AM-3, ESO28SC4   &                23:48:59 & -72:56:43 &    5.5$\pm$0.5  & -1.27          & \cite{dc99} \\
L1, ESO28SC8    &                  0:03:54 & -73:28:19 &    9.0$\pm$1.0  & -1.35$\pm$0.08 & \cite{csp01} \\
L5, ESO28SC16    &                 0:22:40 & -75:04:29 &    4.1          & -1.2$\pm$0.2   & \cite{psg05} \\
K3, L8, ESO28SC19       &          0:24:47 & -72:47:39 &    6.0$\pm$1.3  & -1.16$\pm$0.09 & \cite{csp01} \\
NGC121, K2, L10, ESO50SC12 &       0:26:47 & -71:32:12 &    11.9$\pm$1.3 & -1.71$\pm$0.10 & \cite{csp01} \\
L32, ESO51SC2 &                    0:47:24 & -68:55:10 &    6.7$\pm$0.8  & -1.2$\pm$0.2   & \cite{psc01} \\
L38, ESO51SC3  &                   0:48:50 & -69:52:11 &    5.4$\pm$0.2  & -1.65$\pm$0.2  & \cite{psc01} \\
NGC339, K36, L59, ESO29SC25 &      0:57:42 & -74:28:22 &    6.3$\pm$1.3  & -1.50$\pm$0.14 & \cite{csp01} \\
BS90            &                  0:59:06 & -72:09:03 &    4.3$\pm$0.1  & -0.80          & \cite{ssn07} \\
NGC361, K46, L67, ESO51SC12 &      1:02:11 & -71:36:21 &    8.1$\pm$1.2  & -1.45$\pm$0.11 & \cite{csp01} \\
NGC416, K59, L83, ESO29SC32  &     1:07:59 & -72:21:20 &    6.9$\pm$1.1  & -1.44$\pm$0.12 & \cite{csp01} \\
L110, ESO29SC48              &     1:34:26 & -72:52:28 &   6.4$\pm$1.1   & -1.15$\pm$0.25 & \cite{psg07} \\
L112                         &     1:36:01 & -75:27:28 &   6.7$\pm$1.1   & -1.15$\pm$0.25 & \cite{psg07} \\
BS196                        &     1:48:02 & -70:00:15 &   5.0$\pm$0.5   & -1.68$\pm$0.10  & this paper \\
L113, ESO30SC4               &     1:49:28 & -73:43:42 &   5.3$\pm$1.0   & -1.40$\pm$0.25 & \cite{psg07} \\
\hline
\end{tabular}
\end{minipage}
\end{table*}

In Fig. 9 the angular distribution of the clusters is given overlayed 
on the general population of SMC clusters \citep{b08}. BS\,196 is 
another outer cluster
that fits the SMC axial ratio 1:2 projected on  the sky \citep{csp01}. 
It is the only cluster located 
in a region otherwise clear of known old clusters. The projected 
distribution of old clusters allows a hint on the SMC structure before
the last dynamical interaction with the LMC, which occurred about 200\,Myr
ago \citep{bc07}. Better than an ellipsoidal distribution, the SMC old
clusters follow an exponential-disk profile \citep{b08}.

\begin{figure}
\includegraphics[width=9.5cm]{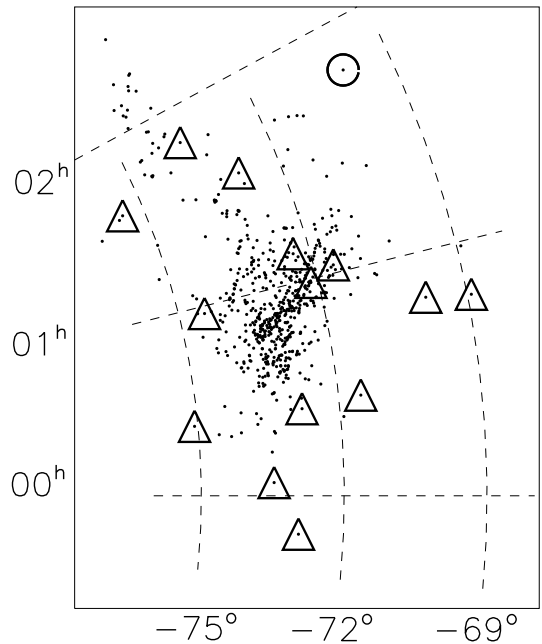}
 \caption{The angular distribution of the old SMC clusters (triangles)
is shown together with the general population of SMC clusters (dots). 
BS\,196 is indicated by a circle.}
\end{figure}

In Fig. 10 we present the SMC initial chemical enrichment determined by 
this population of old clusters. A comparison with the SMC chemical 
enrichment plot by \cite{psg07} shows that BS\,196 behaves like the 
rare metal-poor outlying cluster L\,38 with similar age.
BS\,196 occupies the lower envelope of the distribution in \cite{psg07}. 
A fundamental question is whether lower mass
clusters and/or their spatial distributions provide scatter effects 
in this plot. Larger samples are important to further probe that.

\begin{figure}
\includegraphics[width=8.5cm]{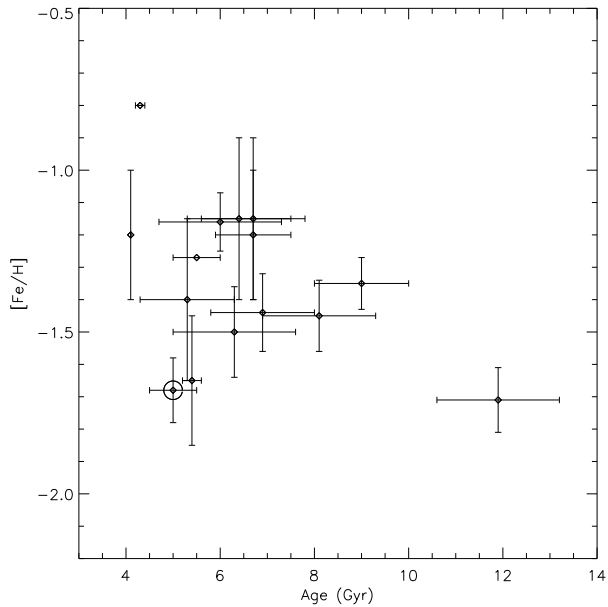}
 \caption{The early chemical enrichment of the SMC adapted from Piatti 
et al. (2007).  BS\,196 is indicated by a circle.}
\end{figure}

\section{CONCLUSIONS}

Since by now most of the populous  old clusters in the SMC have been 
studied, we turned our attention
to fainter clusters in similar environments. We carried out {\it B} and 
{\it V} photometry with the SOAR 4.1-m telescope
of the outlying SMC star cluster BS\,196. It is located in projection
5.1$^\circ$ (5.6\,kpc) from the Northeast of the SMC optical centre, 
thus one of the farthest clusters from the main body.

We confirmed that BS\,196 corresponds to the SMC distance modulus and 
it is thus not a far halo Galactic globular cluster. It is intrinsically 
faint with {\it M}$_{\it V}$=$-1.89\pm0.39$, a new class of objects to be 
explored. We derive an age of 5.0$\pm$0.5\,Gyr and a metallicity
[Fe/H]$=-1.68\pm0.10$. This lower mass cluster is at the lower envelope 
of values in the early chemical enrichment in the SMC.

This might suggest that some outer clusters share this property 
like L\,38 \citep{psc01}.
The cluster structure, fitted by  2 and 3 parameter King profiles, is
well described by a core radius {\it R}$_{\rm c}$=2.66$\pm$0.14\,pc and 
a tidal radius {\it R}$_{\rm t}$=21.2$\pm$1.2\,pc.
The cluster is rather loose 
with a concentration parameter {\it c}=0.90.

The systematic study of such less populous clusters may reveal  
age gap clusters in the LMC, while in the SMC searches for faint additional 
genuine (by age) globular clusters are worth conducting to better 
understand the formation and evolution of star clusters in them. 
The knowledge of the early structure of the Clouds can also benefit
from such studies.

\section*{ACKNOWLEDGEMENTS}
We thank L. Fraga and A. Roman-Lopes for the observations in service 
mode. JFCSJ thanks all SOAR members for making the observatory 
a pleasant environment 
to work during his stay in Chile as resident astronomer.
We thank Dr. A. Piatti for providing information on his recent papers.
This research used the facilities of the Canadian Astronomy Data Centre 
operated by the National Research Council of Canada with the support of 
the Canadian Space Agency. We acknowledge support from the Brazilian 
Institution CNPq.


\begin{thebibliography}{99}

\bibitem[\protect\citeauthoryear{Banks, Dodd \& Sullivan}{1995}]{bds95} 
Banks T., Dodd R. J., Sullivan D. J., 1995, MNRAS, 274, 1225

\bibitem[\protect\citeauthoryear{Bekki \& Chiba}{2007}]{bc07} 
Bekki K., Chiba M., 2007, PASA, 24, 21

\bibitem[\protect\citeauthoryear{Bica \& Bonatto}{2008}]{bb08} 
Bica E., Bonatto C., 2008, MNRAS, 384, 1733

\bibitem[\protect\citeauthoryear{Bica et al.}{2008}]{b08} 
Bica E., Bonatto C., Dutra C. M., Santos Jr. J. F. C., 2008, 
MNRAS, in press, 2008arXiv0806.3049

\bibitem[\protect\citeauthoryear{Bica \& Dutra}{2000}]{bd00} 
Bica E., Dutra C. M., 2000, AJ, 119, 1214

\bibitem[\protect\citeauthoryear{Bica \& Schmitt}{1995}]{bs95}
Bica E., Schmitt H. R., 1995, ApJS, 101, 41 

\bibitem[\protect\citeauthoryear{Burstein \& Heiles}{1982}]{bh82}
Burstein D., Heiles C., 1982, AJ, 87, 1165

\bibitem[\protect\citeauthoryear{Crowl et al.}{2001}]{csp01}
Crowl H. H., Sarajedini A., Piatti A. E., et al., 2001, AJ, 122, 220  
    
\bibitem[\protect\citeauthoryear{Da Costa}{1999}]{dc99}
Da Costa G. S., 1999, in Y.-H. Chu, N. Suntzeff, J. Hesser, D. Bohlender,
eds., IAU Symp. 190, New Views of the Magellanic Clouds, p. 446

\bibitem[\protect\citeauthoryear{Girardi et al.}{2002}]{gbb02}
Girardi L., Bertelli G., Bressan A., et al., 2002, A\&A, 391, 195

\bibitem[\protect\citeauthoryear{Glatt et al.}{2008}]{ggg08}
Glatt K., Gallagher J. S., Grebel E. K. et al., 2008, preprint 
(astro-ph0801.1111)

\bibitem[\protect\citeauthoryear{King}{1962}]{k62}
King I., 1962, AJ, 67, 471

\bibitem[\protect\citeauthoryear{King}{1966}]{k66}
King I. 1966, AJ, 71, 64

\bibitem[\protect\citeauthoryear{Hodge}{1960}]{h60}
Hodge P. W., 1960, ApJ, 131, 351
     
\bibitem[\protect\citeauthoryear{Mighell, Sarajedini \& French}{1998}]{msf98}
Mighell K. J., Sarajedini A., French R. S., 1998, AJ, 116, 2395

\bibitem[\protect\citeauthoryear{Piatti et al.}{2001}]{psc01}
Piatti A. E., Santos Jr. J. F. C., Clari\'a J. J., et al., 
2001, MNRAS, 325, 792

\bibitem[\protect\citeauthoryear{Piatti et al.}{2002}]{psg02}
Piatti A. E., Sarajedini A., Geisler D., Bica E., Clari\'a J. J., 2002, 
MNRAS, 329, 556
  
\bibitem[\protect\citeauthoryear{Piatti et al.}{2005}]{psg05}
Piatti A. E., Sarajedini A., Geisler D. et al., 2005, MNRAS, 358, 1215 

\bibitem[\protect\citeauthoryear{Piatti et al.}{2007}]{psg07}
Piatti A. E., Sarajedini A., Geisler D., Gallart C., Wischnjewsky M., 
2007, MNRAS, 381, L84 

\bibitem[\protect\citeauthoryear{Rich et al.}{2000}]{r00}
Rich R. M., Shara M., Fall S. M., Zurek D., 2000, AJ, 119, 197

\bibitem[\protect\citeauthoryear{Sabbi et al.}{2007}]{ssn07}
Sabbi E., Sirianni M., Nota A. et al., 2007, AJ, 133, 44

\bibitem[\protect\citeauthoryear{Santos, Bica \& Dottori}{1990}]{sbd90}
Santos Jr. J. F. C., Bica E., Dottori H., 1990, PASP, 102, 454 

\bibitem[\protect\citeauthoryear{Schlegel, Finkbeiner \& Davis}{1998}]{sfd98}
Schlegel D., Finkbeiner D., Davis M., 1998, ApJ, 500, 525

\bibitem[\protect\citeauthoryear{Stetson}{2000}]{s00}
Stetson P. B., 2000, PASP, 112, 925 

\bibitem[\protect\citeauthoryear{Tody}{1993}]{t93}
Tody D., 1993, in Hanisch R.J., Brissenden R.J.V., Barnes J., eds.,
IRAF in the Nineties: Astronomical Data Analysis 
Software and Systems II. Astron. Soc. Pac. Conference Ser., 
San Francisco, Vol. 52, 173

\bibitem[\protect\citeauthoryear{Westerlund}{1997}]{w97}
Westerlund B. E., 1997, ``The Magellanic Clouds'', Cambridge Astrophysics
Series \#29, Cambridge Univ. Press.

\bibitem[\protect\citeauthoryear{Yi}{2001}]{y01}
Yi S., Demarque P., Kim Y.-C., et al., 2001, ApJS, 136, 417

\end{thebibliography}
\end{document}